# Spontaneous Hall Effect enhanced by local Ir moments in epitaxial Pr$_2$Ir$_2$O$_7$ thin films

Lu Guo,[1] Neil Campbell,[2] Yongseong Choi,[3] Jong-Woo Kim,[3] Philip J. Ryan,[3,6] Huaixun Huyan,[4] Linze Li,[4] Tianxiang Nan,[1] Jong-Hong Kang,[1] Chris Sundahl,[1] Xiaoqing Pan,[4,5,7] M.S. Rzchowski,[2] Chang-Beom Eom[1*]

[1]Department of Materials Science and Engineering, University of Wisconsin-Madison, Madison, Wisconsin 53706, USA

[2]Department of Physics, University of Wisconsin-Madison, Madison, Wisconsin 53706, U6SA

[3]Advanced Photon Source, Argonne National Laboratory, Argonne, Illinois 60439, USA

[4]Department of Materials Science and Engineering, University of California, Irvine, California 92697, USA

[5]Department of Physics and Astronomy, University of California, Irvine, California 92697, USA

[6]School of Physical Sciences, Dublin City University, Dublin 9, Ireland

[7]Irvine Materials Research Institute, University of California, Irvine, CA 92697, USA

Rare earth pyrochlore Iridates (*RE*$_2$Ir$_2$O$_7$) consist of two interpenetrating cation sublattices, the *RE* with highly-frustrated magnetic moments, and the Iridium with extended conduction orbitals significantly mixed by spin-orbit interactions. The coexistence and coupling of these two sublattices create a landscape for discovery and manipulation of quantum phenomena such as the topological Hall effect, massless conduction bands, and quantum criticality. Thin films allow extended control of the material system via symmetry-lowering effects such as strain. While bulk Pr$_2$Ir$_2$O$_7$ shows a spontaneous hysteretic Hall effect below 1.5K, we observe the effect at elevated temperatures up to 15K in epitaxial thin films on (111) YSZ substrates synthesized via solid phase epitaxy. Similar to the bulk, the lack of observable long-range magnetic order in the thin films points to a topological origin. We use synchrotron-based element-specific x-ray diffraction (XRD) and x-ray magnetic circular dichroism (XMCD) to compare powders and thin films to attribute the spontaneous Hall effect in the films to localization of the Ir moments. We link the thin film Ir local moments to lattice distortions absent in the bulk-like powders. We conclude that the elevated-temperature spontaneous Hall effect is caused by the topological effect originating either from the Ir or Pr sublattice, with interaction strength enhanced by the Ir local moments. This spontaneous Hall effect with weak net moment highlights the effect of vanishingly small lattice distortions as a means to discover topological phenomena in metallic frustrated magnetic materials.

*eom@engr.wisc.edu

Rare-Earth pyrochlore Iridates have been the subject of much research interest as a result of predictions and observations of phenomena such as bulk and edge massless conduction, frustrated magnetism, and metal-insulator transitions[1–5]. The key to the intriguing properties of $RE_2Ir_2O_7$ is the intimate coupling of two disparate sublattices of ionic $RE$ and conducting Ir cations, as well as connecting oxygens (not shown). Each sublattice consists of alternating triangular and Kagome planes that are more easily visualized as forming a corner-sharing tetrahedra network (Fig. 1(a)). The high coordination of the lattice allows significant overlap of Ir orbitals simultaneous with frustrated magnetism of the $RE$ ions.

The $RE$ magnetic exchange interactions are mediated by the conducting Ir bands through the RKKY interaction[4]. As a result of the lattice geometry and antiferromagnetic nearest-neighbor coupling, spins on the $RE$ and Ir sublattices are constrained to point toward or away from the center of one of the adjacent tetrahedra, forming frustrated spin-liquid correlations as examples are shown in Fig. 1(b)-(c). The $RE$ ions have typical local moments of a few Bohr magnetons[4], while the Ir ions have a smaller moment which can be local or delocalized[6,7]. Ir adds to the complexity due to its strong spin-orbit coupling, mixing together the orbital and spin degrees of freedom.

In the $RE_2Ir_2O_7$ family, bulk $Pr_2Ir_2O_7$ is unique in that it is metallic down to the lowest temperatures and shows topological Hall effect below 1.5K as a result of various two-in-two-out magnetic configurations at the Pr sites with no net magnetic moment or long-range order. Furthermore, in thin films this topological Hall effect was observed at elevated temperatures up to 50K[8]. While there has been speculation about the role of Ir in this effect[8], here we combine Hall measurements, synchrotron X-ray diffraction and spectroscopy techniques to provide direct evidence supporting the emergence of Ir local moments, induced by the vanishingly small lattice modification of the Ir sublattice in the epitaxial $Pr_2Ir_2O_7$ thin films. We show a definitive link between the emergence of Ir local moments, which are absent in bulk $Pr_2Ir_2O_7$, and the increased onset temperature of the spontaneous Hall effect in the films.

To study the Hall effect in $Pr_2Ir_2O_7$ thin films, we synthesized stoichiometric epitaxial relaxed films via the solid-phase-epitaxy method, see supplemental information for details[9]. The high resolution Scanning Transmission Electron Microscopy (STEM) image across the interface between the film and substrate, as shown in Fig. 2(e), together with X-ray diffraction in Fig. 2(a)-(d), confirms a good epitaxial relationship and a sharp interface. Moreover, the modulated intensity contrast in Fig. 2(e) arises from an atomic number modulation between columns, indicating an atomic arrangement in the film that matches the ordered pyrochlore lattice. Energy Dispersive x-ray Spectroscopy (EDS) confirms the cation ratio between Pr and Ir is almost 1:1 (Fig. S1(e)). While the pyrochlore structure is nominally cubic, the synchrotron X-ray diffraction study shows different d-spacing for the (6 0 10) and (0 $\bar{6}$ 10) reflections, which are equivalent under cubic symmetry. This points toward a breaking of the cubic symmetry in our epitaxial $Pr_2Ir_2O_7$ thin film; however, the distortion is too small for us to discern the specific symmetry of the lattice from the X-ray diffraction study (see supplemental Table S1[9]). There are indications that a trigonal distortion of the Ir sublattice can change the electronic and magnetic properties of $Pr_2Ir_2O_7$[10].

However our result pushes the lower limit of the lattice distortion necessary to effectively alter the Ir local electronic environment enough to produce the spontaneous Hall enhancement.

We use Hall measurements to study the electronic and magnetic manifestations of minor lattice distortions in the epitaxial $Pr_2Ir_2O_7$ films. Fig. 3(b) shows the Hall signal at different temperatures with applied magnetic field along the [111] direction. At temperatures below 20K, the Hall conductivity is non-linear, becomes hysteretic, and develops a small remnant value at zero field, referred to here as the spontaneous Hall effect. A similar effect is observed in the bulk single crystal, but only at temperatures an order of magnitude lower[3,8]. The origin of such effect most commonly occurs from a spontaneous net magnetic moment via the anomalous Hall effect. We rule of this contribution based on our x-ray measurements, which indicate the net Ir moment is less than 0.05μB/Ir at 5 T and the film lacks long-range magnetic ordering (Fig. S2). Consequently, we conclude that the spontaneous Hall effect in the film arises from the topological Hall effect. In this case, as discussed in the supplementary information[9], the time-reversal symmetry is broken from the frustrated spin-liquid correlations rather than a net magnetic moment.

Since the *RE* and Ir cations are both magnetically active, we use element-resolved x-ray magnetic scattering and spectroscopy to explore the individual Pr and Ir sublattice contributions to the spontaneous Hall effect. X-ray resonant diffraction measurements at the Ir $L_3$ edge of our thin films from 65K down to 5K, covering the temperature regime above and below the observed onset of the spontaneous Hall effect, show no clear indications of any type of long-range magnetic ordering (Fig. S7-S8) including Ir-site AIAO ordering, consistent with the intrinsic bulk single-crystal behavior.

To help us understand the $Pr_2Ir_2O_7$, we compare the Pr $L_2$-XMCD results of our thin film with that of cubic-symmetric $Pr_2Ir_2O_7$ powders. Comparison of film Pr $L_2$-XMCD peaks (Fig. 4) and the reference powder peaks (Fig. S3) suggests that Pr spins behave identically in films and powders. In addition, the field dependence of the Pr $L_2$-XMCD signal from the film (Fig. S4(a)) resembles the magnetometry result from the powder sample (Fig. S2(a)). These results suggest that the structural distortion in the film does not modify the Pr magnetism. This makes sense as the Pr 4*f* orbitals are fairly localized and overlap less with the distorted surrounding atoms. This contrasts sharply with the Ir 5*d* orbitals that are much more dispersive and so more susceptible to structural distortion effects. Previous studies have shown that distortion of the $IrO_6$ octahedra in perovskite iridate superlattices induces long-range magnetic ordering and an insulating ground state[11].

To investigate potential changes in the Ir magnetism in the films, we compared the Ir $L_{2,3}$ XAS and XMCD data from the $Pr_2Ir_2O_7$ film with $Pr_2Ir_2O_7$ and $Sr_2IrO_4$ powders. As shown in Fig. 5 and supplemental Fig. S5, the absorption edge positions and the $L_3/L_2$ branching ratios are similar for both powders, as well as the film, and also with previously studied $SrIrO_3/La_{0.3}Sr_{0.7}MnO_3$[7], confirming large spin-orbit coupling and a $Ir^{4+}$ electronic environment in all materials[12]. Despite the similarity of the Ir orbital state, the XMCD signal under 5T on $Pr_2Ir_2O_7$ powders at the Ir $L_3$ edge is less than 10% of that in $Sr_2IrO_4$ powders. The observed weak Ir-XMCD value is expected due to the lack of long-range ordering on the Ir 5*d* moments[3]. Furthermore, the Ir-XMCD results here differ significantly from the cases related to spin-orbit coupled Ir local moments with strong orbital magnetic moments[6,7]. Unexpectedly, the Ir-$L_3$ XMCD sign from the $Pr_2Ir_2O_7$ powder indicates that the Ir induced moments are anti-parallel to the external field, in sharp contrast to the $Sr_2IrO_4$ case, which shows long-range ordered moments with a net moment of 0.05 $\mu_B$/Ir[6]. Disparate Ir-$L_2$ XMCD signals seen in these powder samples provide clues to understand the discrepancy in the Ir-$L_3$ XMCD signals and the overall magnetism in these two materials. Whereas the negligible Ir-$L_2$ XMCD signal (in comparison with the $L_3$ signal) from the $Sr_2IrO_4$ is the characteristic signature of the $J_{eff}$=1/2 state, the $Pr_2Ir_2O_7$ powders show comparable amplitudes with opposite signs between the Ir $L_2$ and $L_3$ edges. This reveals that the observed Ir spin moments in the $Pr_2Ir_2O_7$ powder samples originate from small spin polarization in the conduction band by Pr moments, distinctively different from localized Ir moments in the $Sr_2IrO_4$ case. This is in accordance with resistivity results which suggest small conduction electron magnetization via the Kondo effect as shown in Fig. 3(a).

We now discuss the difference between $Pr_2Ir_2O_7$ film and powders, in particular that our XMCD data show a magnetically active Ir sublattice in the film, as opposed to the weak polarization of the Ir spins we observed in powders. The Ir-XMCD signals in the film are an order of magnitude larger than in the $Pr_2Ir_2O_7$ powder, indicating a field-induced alignment of local Ir moments, as shown in Fig. 5. The sign of Ir-$L_3$ XMCD in the $Pr_2Ir_2O_7$ thin film, contrary to that of the $Pr_2Ir_2O_7$ powders, indicates that the net Ir moment aligns with the external field, and thus is parallel to the Pr 4*f* net moment. Moreover, the much smaller XMCD signal at the $L_2$ edge relative to $L_3$ suggests that, unlike the $Pr_2Ir_2O_7$ powder, the Ir 5*d* electronic state in the $Pr_2Ir_2O_7$ thin film is close to the $J_{eff}$ = 1/2 state which is observed in $Sr_2IrO_4$[6,7]. We thus conclude that the vanishingly-small structural distortion present in our thin films has a profound impact on the electronic properties by localizing the Ir moments and changing the Ir $t_{2g}$ manifold toward the $J_{eff}$ = 1/2 observed in other iridates[13].

Our findings are also consistent with the theoretical predictions that a trigonal distortion of the Ir sublattice can enhance the formation of Ir local moments[14]. Previous theoretical work suggested a trigonal distortion of the Ir sublattice can affect the octahedral crystal field and result in non-trivial topological phases in $RE_2Ir_2O_7$ by stabilizing the AIAO spin configuration at Ir sites[10]. Our results highlight the disproportionate effect of a trivial lattice distortion on the Ir sublattice inducing incipient Ir 5d local moments as a manifestation of perturbations to the $t_{2g}$ bands[15].

Previous work notes that the Pr-Pr interaction strength is too weak to allow chiral spin liquid correlations on the Pr sublattice at temperatures above 1.5K, despite measurable spontaneous Hall effect signal. The origin of the observed effect was attributed to broken time reversal symmetry at the Ir sublattice[8]. As the most similar point of comparison to $Pr_2Ir_2O_7$ is $Nd_2Ir_2O_7$, which shows AIAO spin ordering at both the Nd and Ir sites in the insulating phase[16], Ohtsuki et al propose AIAO to be the most likely configuration for Ir in $Pr_2Ir_2O_7$[8]. The AIAO ordering and its relationship to the A site elements have been well linked throughout the rare earth pyrochlore iridate series (apart from Pr), with neutron scattering and resonant elastic and inelastic x-ray scattering[16–19]. But we did not observe any signal that would indicate long-range AIAO ordering in the Ir sublattice (Fig. S7-S8).

As a result of not observing Ir-AIAO ordering, we consider other ways in which Ir moments play a role in the observed TRS breaking. One possible mechanism is that the Ir local moments, having the same frustrated lattice structure as the Pr, form the same chiral spin liquid correlations, just at higher temperatures. The spontaneous Hall effect in bulk $Pr_2Ir_2O_7$ has been attributed to altering dominant 3-in/out-1-out/in and 2-in-2-out correlations variants at the Pr sites, which leads us to presume these same correlations at the Ir sites can produce the spontaneous Hall effect. This is consistent with the Ir-site spin correlations preceding the Pr-site correlations with cooling, as observed in $RE_2Ir_2O_7$ materials through their MIT. In this scenario, the observed small Ir net moment, presumably from canting or defects, couples with the external field, allowing the external field to manipulate the Ir-site spin liquid correlations while not directly producing the anomalous Hall effect[20].

A second possible mechanism is that the Ir local moments renormalize the effective Pr-Pr interaction strength, raising the temperature of the spin-liquid correlation onset of the Pr moments. Since the RKKY interaction describes the Pr-Pr coupling as mediated by the Ir conduction band, consistent with AFM coupling and nonzero conductivity, spin polarization at the Ir site should impact this interaction. Subsequently, on the basis of drastically different conduction properties of the Pr and Ir sublattices, one would expect Ir-site spin-liquid correlations to manifest differently in the spontaneous Hall effect than the Pr-site correlations. Yet despite this, the spontaneous Hall effect hysteresis loops appear very similar to the bulk loops attributed to Pr-site correlations[3]. In this scenario, the Ir sublattice would maintain no chiral spin-liquid correlations at all temperatures, and the Ir TRS breaking would only indirectly contribute to the spontaneous Hall effect. The Pr sublattice chiral spin liquid correlations would be the 2-in-2-out and 3-in/out-1-out/in correlations responsible for the spontaneous Hall effect in the bulk[3]. In $Eu_2Ir_2O_7$ films (with no magnetic contribution from $Eu^{3+}$ ions), anomalous Hall effect has been observed, highlighting the contribution of magnetic Ir ions[21]. At this point our experimental evidence does not rule out either scenario, and future studies on the $RE_2Ir_2O_7$ (without $RE$ local moment) may shed further light on the role of the magnetic Ir sublattice in the Hall effect.

In conclusion, we observed that small lattice distortions in $Pr_2Ir_2O_7$ thin films act as a perturbation, changing the local magnetic properties of the Ir sublattice and inducing a spontaneous Hall effect at elevated temperatures. By analyzing the XMCD signal at the Pr and Ir $L$ edges, we attribute the enhanced spontaneous Hall response in the thin films to localized net moments on the Ir sublattice. Our observations reveal the possible link between structural change in the Ir network, the Ir local magnetic environment, and transport behavior. Understanding these effects provides opportunities to manipulate the $5d$ pyrochlore iridate ground states by modifying the lattices, making this system attractive as a promising candidate in spintronics. Our work opens up new possibilities for controlling electronic and magnetic phenomena in conducting frustrated antiferromagnets via thin-film epitaxy.

**Acknowledgements**

Synthesis of thin films at the University of Wisconsin-Madison was supported by NSF through the University of Wisconsin Materials Research Science and Engineering Center (DMR-1720415) and DMREF (Grant No. DMR-1629270). Transport and magnetic measurements at the University of Wisconsin-Madison was supported by the US Department of Energy (DOE), Office of Science, Office of Basic Energy Sciences, under award number DEFG02-06ER46327. X-ray diffraction and absorption experiments were carried out at beamlines 6-ID-B and 4-ID-D of the Advanced Photon Source, Argonne National Laboratory. The work performed at the Advanced Photon Source was supported by the U.S. Department of Energy, Office of Science, and Office of Basic Energy Sciences under Contract No. DEAC02-06CH11357. TEM experiments were conducted using the advanced TEM facilities in the Irvine Materials Research Institute (IMRI) at the University of California, Irvine.

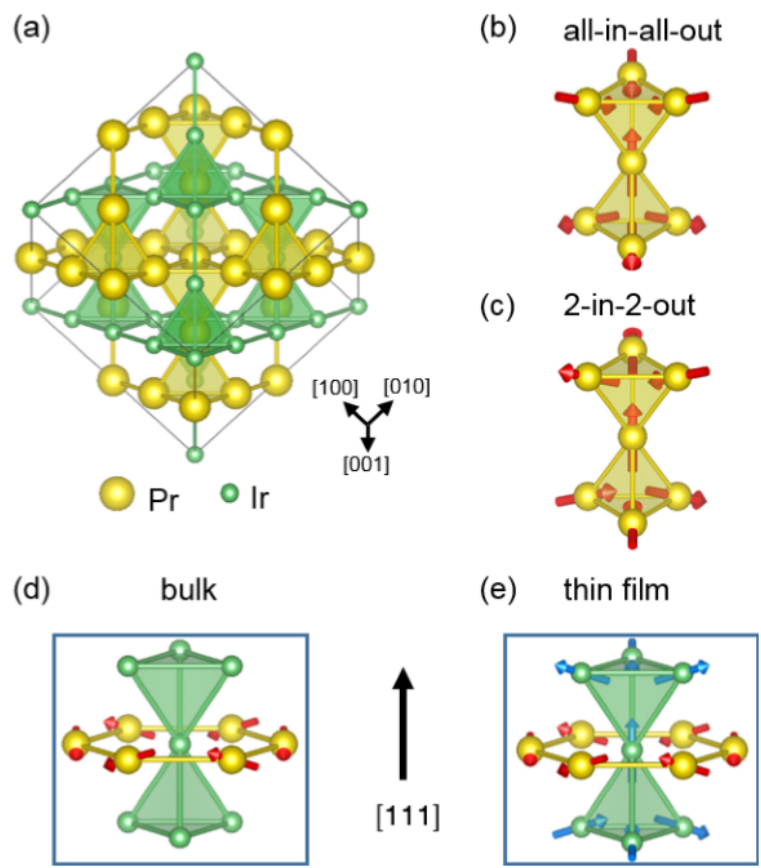

**Figure 1:** (a) Unit cell of $Pr_2Ir_2O_7$ with only cation sublattices shown. (b, c) 'All-in-all-out' and '2-in-2-out' spin configurations on the Pr corner sharing tetrahedra. (d, e) Zoom-in of corner-sharing Ir tetrahedral surrounded by Pr hexagonal ring in the (111) plane of the bulk and thin film respectively. The red arrows in the Pr atoms indicate the Pr 4f moments. Due to the cubic symmetry-breaking in the thin film, Ir local moments can be established indicated here in (e) by the blue arrows in Ir atoms.

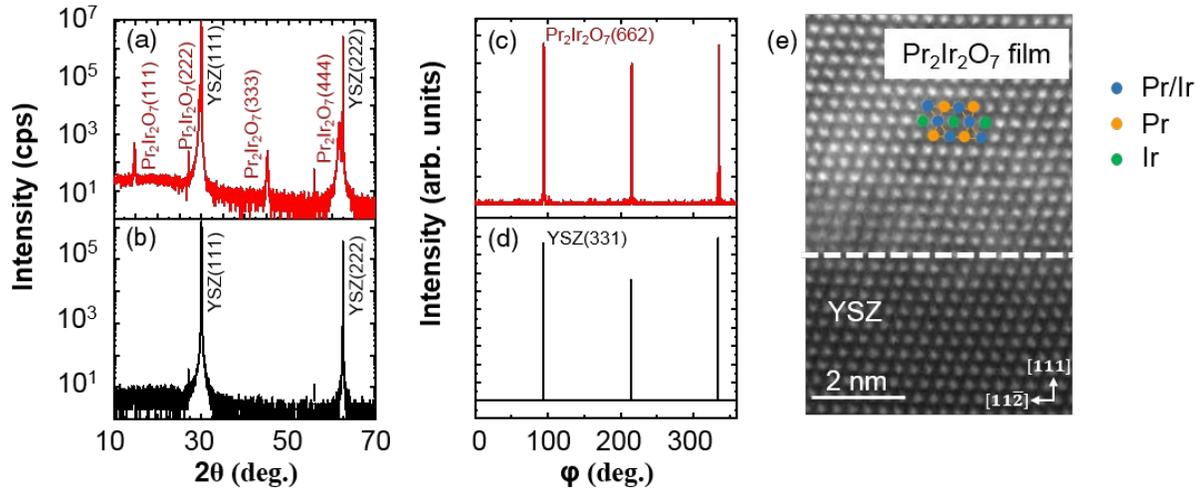

**Figure 2:** Out-of-plane 2θ-ω scan of (a) the post-annealed and (b) the as-grown film. Phi-scan patterns of the (c) {662} planes from the epitaxial crystalline $Pr_2Ir_2O_7$ thin film and (d) {331} planes from the YSZ (111) substrate. (e) Cross-sectional HRSTEM image across the interface between epitaxial $Pr_2Ir_2O_7$ thin film and (111) YSZ substrate. The Pr, and Ir atomic positions labeled in the selected areas are consistent with pyrochlore structure. The blue dots represent the mixed Pr and Ir column due to the alternating arrangement along zone axis. The O atoms are omitted.

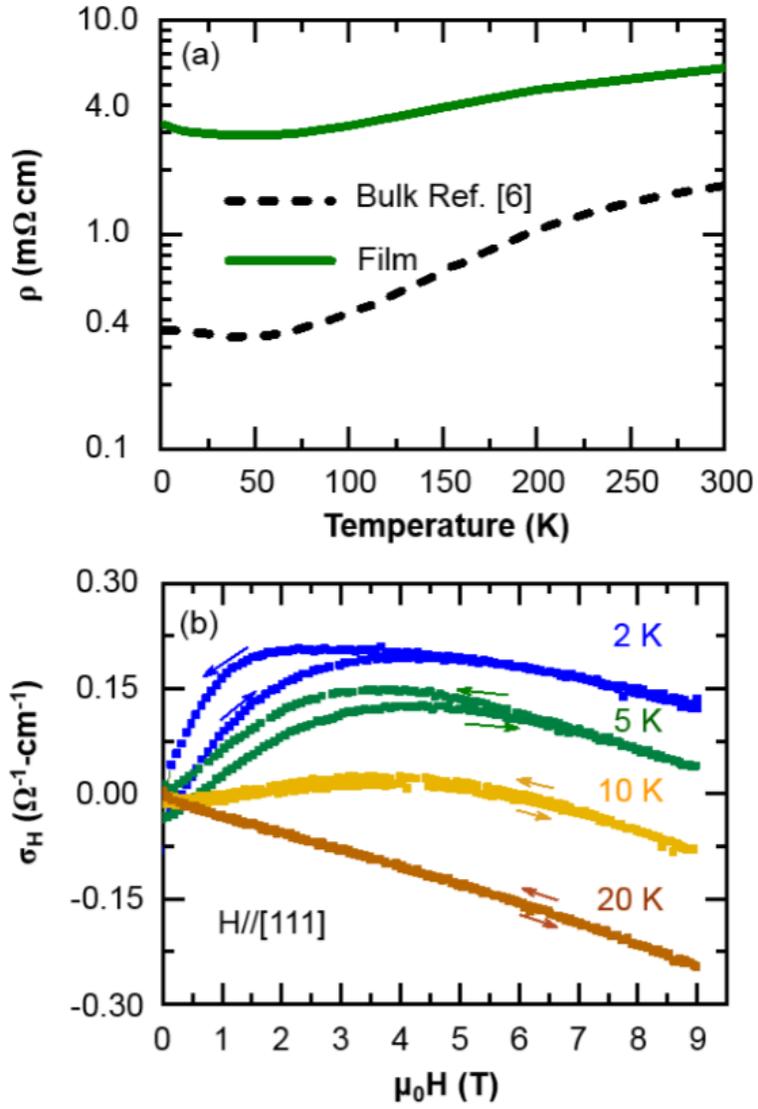

**Figure 3:** (a) Temperature dependence of the longitudinal resistivity of the epitaxial $Pr_2Ir_2O_7$ thin film (solid green line) and the bulk $Pr_2Ir_2O_7$ single crystal (black dashed line) from Ref. 6. (b) Hall conductivity as a function of external out-of-plane magnetic field at different temperatures.

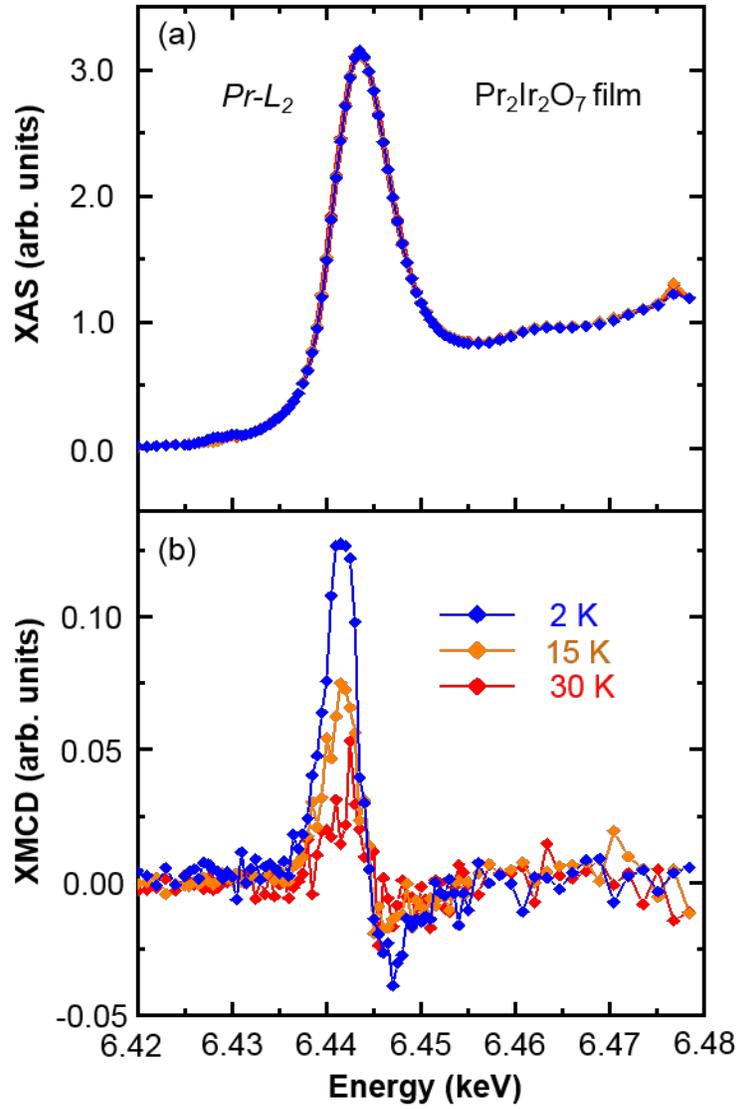

**Figure 4:** (a) XAS and (b) XMCD spectra at the Pr $L_2$ edge under 5T magnetic field on $Pr_2Ir_2O_7$ thin films through the spontaneous Hall effect transition.

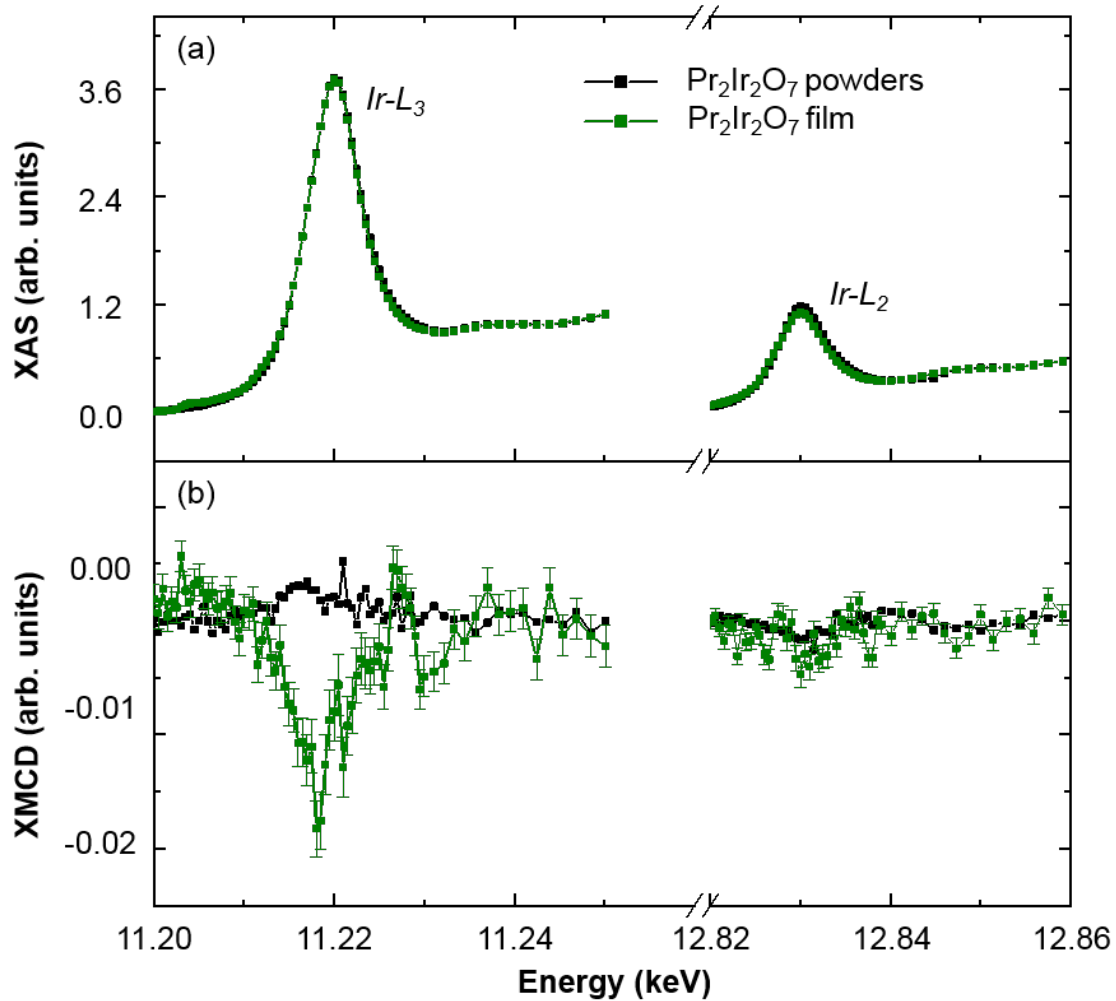

**Figure 5:** The XAS and the XMCD spectra at 2K under 5T with magnetic field along the [111] direction at the Ir $L_{2,3}$ edges are compared between the $Pr_2Ir_2O_7$ film (green) and $Pr_2Ir_2O_7$ reference powders (black). The main difference appears in the Ir-$L_3$ XMCD signals, in which the thin film XMCD signal shows the opposite sign and larger magnitude compared to the reference powders.

# Supplementary Information for

## Spontaneous Hall Effect enhanced by local Ir moments in epitaxial $Pr_2Ir_2O_7$ thin films

Lu Guo,[1] Neil Campbell,[2] Yongseong Choi,[3] Jong-Woo Kim,[3] Philip J. Ryan,[3,6] Huaixun Huyan,[4] Linze Li,[4] Tianxiang Nan,[1] Jong-Hong Kang,[1] Chris Sundahl,[1] Xiaoqing Pan,[4,5,7] M.S. Rzchowski,[2] Chang-Beom Eom[1*]

[1]Department of Materials Science and Engineering, University of Wisconsin-Madison, Madison, Wisconsin 53706, USA

[2]Department of Physics, University of Wisconsin-Madison, Madison, Wisconsin 53706, U6SA

[3]Advanced Photon Source, Argonne National Laboratory, Argonne, Illinois 60439, USA

[4]Department of Materials Science and Engineering, University of California, Irvine, California 92697, USA

[5]Department of Physics and Astronomy, University of California, Irvine, California 92697, USA

[6]School of Physical Sciences, Dublin City University, Dublin 9, Ireland

[7]Irvine Materials Research Institute, University of California, Irvine, CA 92697, USA

*eom@engr.wisc.edu

1. **Synthesis of the $Pr_2Ir_2O_7$ thin film by solid phase epitaxy method**

Amorphous $Pr_2Ir_2O_7$ thin films were deposited on the Y-stabilized (YSZ) (111) single-crystal substrates via on-axis RF sputtering at room temperature under 15mTorr of 90%Ar and 10%$O_2$ mixed gas. Polycrystalline $Pr_2Ir_2O_7$ target was used for deposition. After deposition, crystalline $Pr_2Ir_2O_7$ thin films were formed during the following annealing process in the electronic furnace at $800^0$C for 12 hours in air. To study the microstructure of the epitaxial $Pr_2Ir_2O_7$ thin film, Scanning Transmission Electron Microscopy (STEM) was performed. In order to study the stoichiometry information, Energy Dispersive x-ray Spectroscopy (EDS) mapping was performed to study the elemental distribution. The atomic ratio between Pr and Ir is calculated to be close to 1:1 by integrating the signal from each element within the area, demonstrating that the crystalline $Pr_2Ir_2O_7$ thin film is close to stoichiometric, as shown in supplementary Fig. S1(f). This is supported by our longitudinal resistivity measurements, shown in Fig. 2(a). Comparing to the bulk, the $Pr_2Ir_2O_7$ thin films show similar metallic temperature dependence with an upturn in resistivity below 50K. This similarity buttresses our claim that the film is stoichiometric, as off-stoichiometric films are known to show quite different transport behavior[27].

## 2. Structure analysis of the Pr$_2$Ir$_2$O$_7$ thin film by XRD, AFM, and STEM

The reciprocal space map around YSZ (331) and Pr$_2$Ir$_2$O$_7$ (662) reflection peaks reveals that the film is fully relaxed, as shown in Fig. S1(a). Atomic Force Microscopic (AFM) image reveals the film surface becomes rougher after post-annealing process as shown in Fig. S1(b) and Fig. S1(c). Fig. S1(d) shows low magnification STEM cross-section images. The island nature of the Pr$_2$Ir$_2$O$_7$ thin film is observed, possibly due to the wetting issue between the film and the substrate. The size of the crystalline islands is on the order of hundred nms. This feature is consistent with the measured rougher surface after annealing in AFM image in Fig. S1(c). EDS mapping results across the interface between crystalline Pr$_2$Ir$_2$O$_7$ thin film and YSZ substrate are shown in Fig. S1(e). Good epitaxial relationship can be observed. The left column displays the dark field TEM image, in which the bright and dark parts represent Pr$_2$Ir$_2$O$_7$ thin film and YSZ substrate, respectively. The middle and right column display the element mapping across the interface for Pr (blue) and Ir (magenta), respectively.

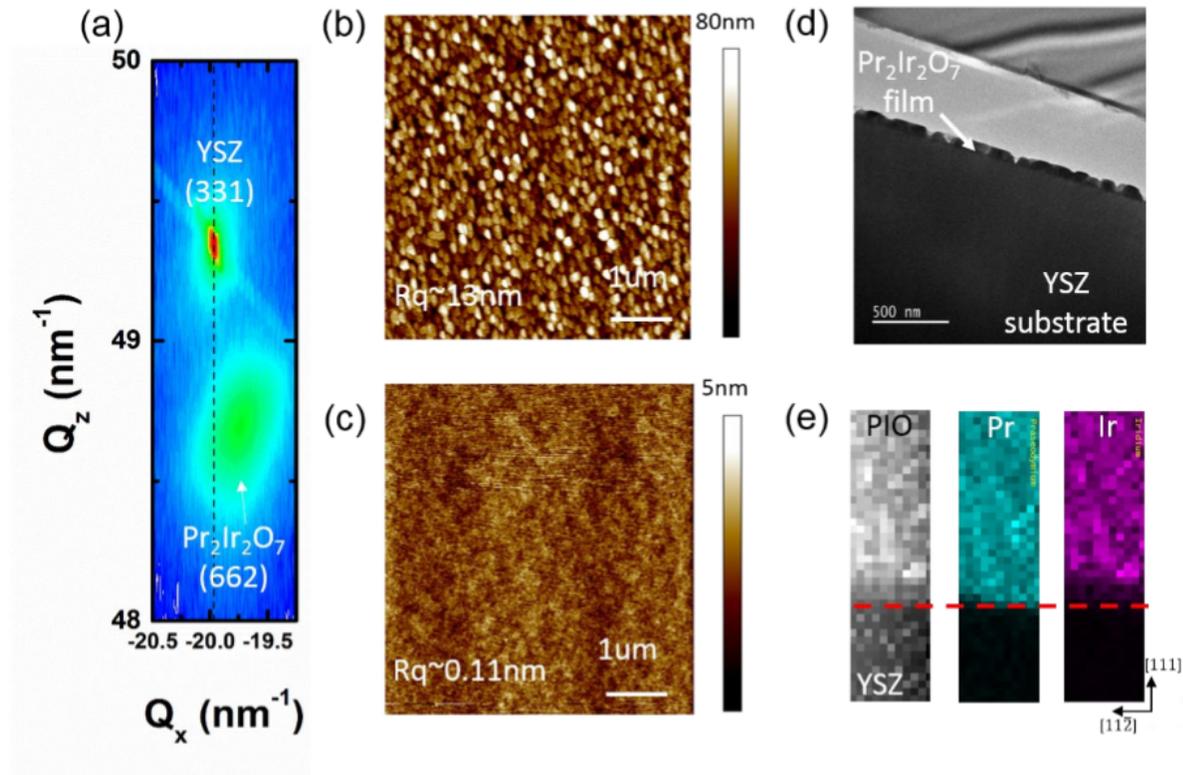

FIG. S1: (a) Reciprocal space map around YSZ (331) and Pr$_2$Ir$_2$O$_7$ (662) reflection peaks. Atomic Force Microscopy topology results of (b) the post annealed epitaxial crystalline Pr$_2$Ir$_2$O$_7$ thin film and (c) the as-grown amorphous Pr$_2$Ir$_2$O$_7$ thin film. (d) Low magnification STEM cross section images. (e) EDS mapping across the interface between crystalline Pr$_2$Ir$_2$O$_7$ thin film and YSZ substrate. The red dotted line indicates the location of interface based on EDS.

## 3. Rhombohedral distortion in the Pr$_2$Ir$_2$O$_7$ thin film

To verify the lattice symmetry of Pr$_2$Ir$_2$O$_7$ thin films, X-ray diffraction measurements were performed on out-of-plane and in-plane reflection peaks, as shown in Table S1. At room temperature, the measured in-plane peaks (6 0 10) and (0 $\bar{6}$ 10) show different 2θ values, suggesting the broken cubic symmetry in the thin film. Considering the in-plane symmetry of (111) orientation, tetragonal distortion, monoclinic distortion, hexagonal distortion and triclinic distortion can be ruled out due to incompatible symmetry. Between trigonal and orthorhombic distortion, the measured (6 0 10) and (0 $\bar{6}$ 10) planes suggest to fit trigonal distortion better though the precise symmetry analysis is beyond experimental resolution of our data.

| h | k | l | 2θ (measured) | 2θ (cubic) | 2θ (rhombohedral) | 2θ (orthorhombic) |
|---|---|---|---|---|---|---|
| 1 | 1 | 1 | 7.8758 | 7.8821 | 7.8773 | 7.8294 |
| 2 | 2 | 2 | 15.7740 | 15.8018 | 15.7922 | 15.6960 |
| -2 | -2 | 10 | 48.7083 | 48.7089 | 48.7259 | 48.9400 |
| -1 | -1 | 11 | 52.2288 | 52.2188 | 52.2310 | 52.5300 |
| 0 | 0 | 12 | 56.8780 | 56.8718 | 56.8778 | 57.2350 |
| 1 | 1 | 13 | 62.5296 | 62.5170 | 62.5161 | 62.909 |
| 2 | 2 | 14 | 69.0415 | 69.0495 | 69.0409 | 69.453 |
| 6 | 0 | 10 | 55.1067 | 55.1306 | 55.1180 | 55.162 |
| 0 | -6 | 10 | 55.1506 | 55.1306 | 55.1548 | 55.206 |

Table S1: 2θ values of out-of-plane and in-plane reflection peaks of the epitaxial Pr$_2$Ir$_2$O$_7$ thin film. Measured values are from synchrotron X-ray diffraction at room temperature with incident X-ray energy of 15keV. By fitting the measured values with different symmetry, the corresponding 2θ values are tabulated and listed as above.

## 4. Topological Hall Effect from Spin-Liquid Correlations

As discussed in the paper, the spontaneous Hall effect in the film likely arises from the topological Hall effect. Though in Pr$_2$Ir$_2$O$_7$ the observed spontaneous Hall effect implies time-reversal symmetry breaking, there is no evidence of long-range magnetic order indicating the role of spin-liquid correlations. Topological Hall effect results from a non-zero macroscopic average of the scalar spin chirality arising from spin-liquid correlations, defined as $<\Omega_{ijk}> = <\vec{S}_i \cdot \vec{S}_j \times \vec{S}_k>$, where $i, j$, and $k$ are any 3 neighboring spins. Such a configuration of local spins is known to produce a spontaneous Hall effect in conducting materials[1]. This Hall effect has its origin in the Berry phase accumulated by an electron moving through the lattice whose spin is rotated adiabatically through the spin configuration and returns to its initial orientation[1,2]. This Berry phase accumulation is identical to how the electron phase would be rotated by the vector potential of an out-of-plane constant magnetic field. The scalar spin chirality is proportional to the phase rotation, allowing us to link the spontaneous Hall effect to the magnetic spin-liquid correlations. Single crystals of Pr$_2$Ir$_2$O$_7$ only show the spontaneous Hall effect below 1.5K, but our observation at much higher temperature necessitates further understanding of the magnetism.

## 5. Magnetometry result on $Pr_2Ir_2O_7$ powders and $Pr_2Ir_2O_7$ thin film:

$Pr_2Ir_2O_7$ powders are made by mixing $Pr_6O_{11}$ powders and $IrO_2$ powders, and heating the mixed powders in air at 1000C for 3 days with several interval grindings. Magnetic moments of $Pr_2Ir_2O_7$ powders are measured via superconducting quantum interference device (SQUID). The magnetic field is applied between -9T and 9T. Fig. S2(a) shows the field dependent magnetic moment of $Pr_2Ir_2O_7$ powders at different temperatures. The non-linear M vs B curve without hysteresis between field sweeping up and sweeping down resemble the bulk behavior and can be expected from the non-collinear spin textures on Pr-sublattices. Magnetic moments of $Pr_2Ir_2O_7$ thin film is also measured, as shown in Fig. S2(b). No observation of measurable hysteresis in M vs T curve in the film down to 2K. The signal in Fig. S2(b) mostly comes from the YSZ bare substrate, without showing net magnetization from $Pr_2Ir_2O_7$ thin film. Combining the magnetic x-ray scattering study on Pr and Ir sublattices, no long-range ordering magnetism is established in the film, or responsible for the observed SHE.

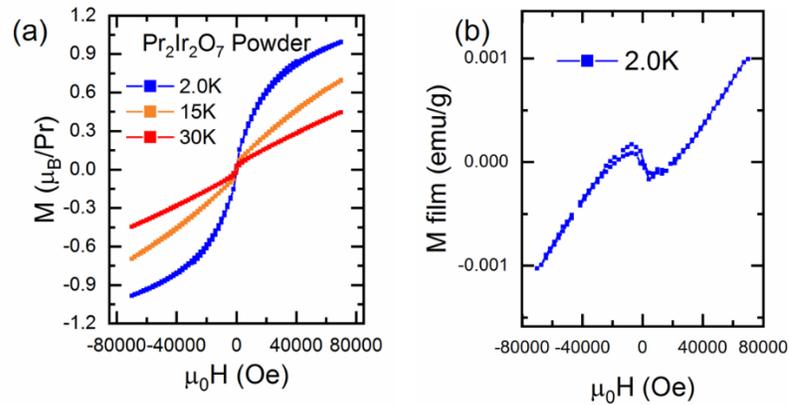

FIG. S2: (a) M vs B curve measured by SQUID on $Pr_2Ir_2O_7$ powders at different temperatures. The inset labels describe the measurement temperature. (b) M vs B curve measured by SQUID on $Pr_2Ir_2O_7$ thin film at 2K. The signal mostly comes from the substrate rather than the $Pr_2Ir_2O_7$ film.

## 6. Temperature dependent Pr-XMCD and Ir-XMCD on Pr₂Ir₂O₇ powders under 5T:

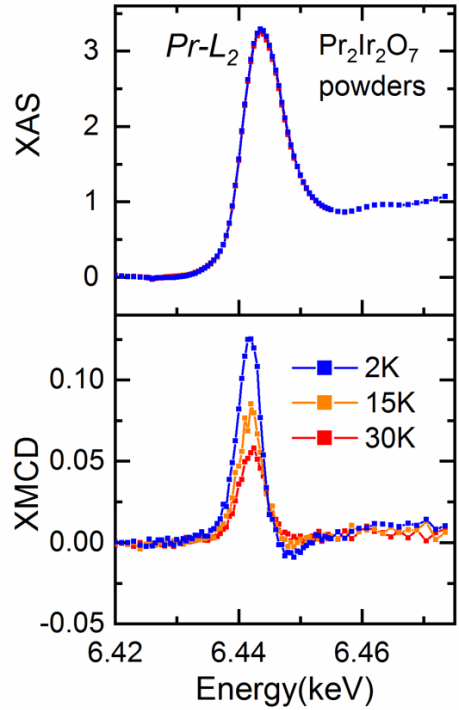

FIG. S3: The temperature dependent XAS(the solid lines) and XMCD(the solid circles) at the Pr $L_2$ edge under 5T on the Pr$_2$Ir$_2$O$_7$ powders at different temperatures.

## 7. Temperature dependent Pr-XMCD and Ir-XMCD on Pr₂Ir₂O₇ thin film with applying in-plane magnetic field:

The asymmetric ratio from XMCD signal at the Pr $L_2$ edge from 2K to 40K under 5T is shown in Fig. S4(a). The temperature dependence resembles the magnetization behavior of Pr$_2$Ir$_2$O$_7$ bulk. The Ir-XMCD of the Pr$_2$Ir$_2$O$_7$ film at the Ir L edge under 5T at 2K are shown in Fig. S4(b) and Fig. S4(c). The opposite sign of Ir-XMCD in the Pr$_2$Ir$_2$O$_7$ thin film at the $L_3$ edge from that of the powders with magnetic field parallel to [1$\bar{1}$0] (in-plane) direction, as shown in Fig.4(b), also indicates that the net Ir moments follow the external field to align, and are parallel to the Pr 4f moments, which is consistent with the results with applying out-of-plane magnetic field.

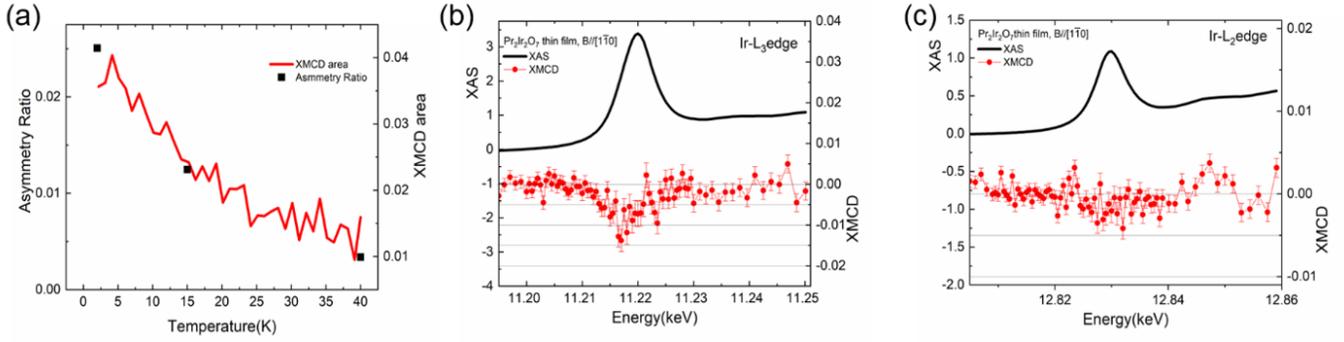

FIG. S4: XAS and XMCD on the $Pr_2Ir_2O_7$ film. (a) The asymmetric ratio from XMCD signal at the Pr $L_2$ edge from 2K to 40K under 5T. (b) ,(c) The XAS (the black solid lines) and the XMCD at 2K under 5T (the red solid circles) with magnetic field along $[1\bar{1}0]$ at the Ir $L_3$ ,and $L_2$ edge, respectively.

## 8. XAS and XMCD at the Ir edge on $Pr_2Ir_2O_7$ powders with magnetic field along [111]:

We performed XAS and XMCD at the Ir edge on $Pr_2Ir_2O_7$ powders are performed at 5T and 2K, as shown in supplementary Fig. S5. The Ir $L_3$ XMCD sign from the powder indicates the Ir net moments are anti-aligned with the external field, in sharp contrast to the $Sr_2IrO_4$ case. Whereas the negligible Ir $L_2$ XMCD signal (in comparison with the $L_3$ signal) from the $Sr_2IrO_4$ is the characteristic signature of the $J_{eff}=1/2$ state, the $Pr_2Ir_2O_7$ powders show comparable amplitudes with opposite signs between the Ir $L_2$ and $L_3$ edges.

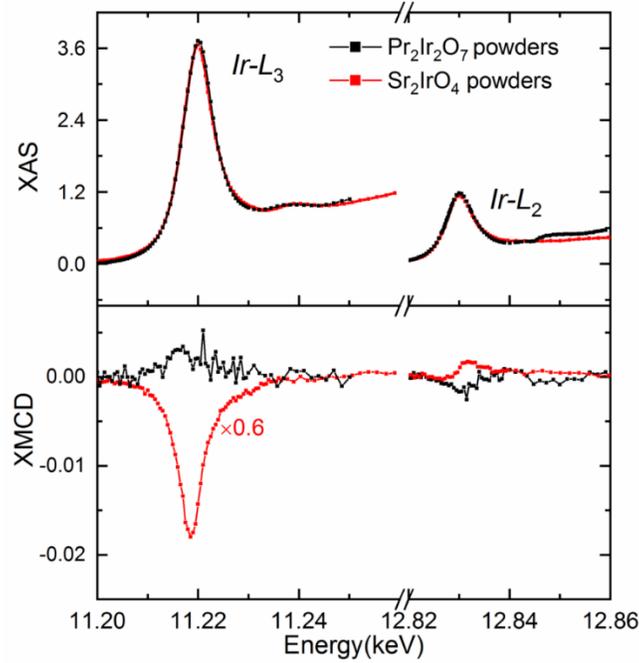

FIG. S5: The XAS and the XMCD at 2K under 5T with magnetic field along [111] at the Ir $L_{2,3}$ edge on $Pr_2Ir_2O_7$ powders. For comparison, XAS and XMCD (the red solid lines which is scaled by multiple 0.6 for direct value comparison) data at the Ir $L_3$ edge on $Sr_2IrO_4$ powers from Ref. 30 are included.

9. **Contour color map of the anomalous hall conductivity of $Pr_2Ir_2O_7$ thin film in the T-B plane:**

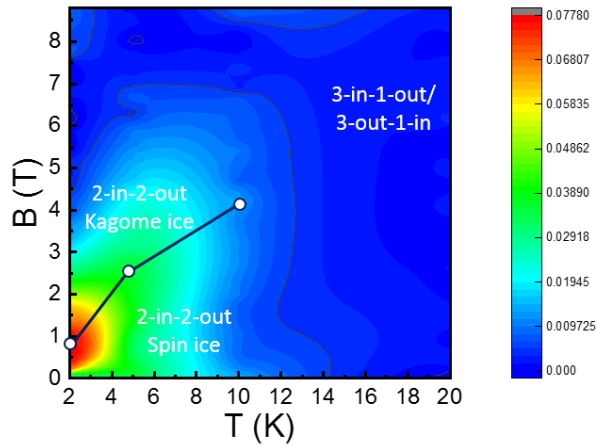

FIG. S6: Color map of anomalous Hall conductivity in the T-B plane. According to the field dependence of Hall conductivity, 2-in-2-out, and 3-in-1-out/3-out-1-in spin structures on Pr sub-lattices stabilization regimes are indicated. The blue empty circles represent the transition field, corresponding to the field with maximum Spontaneous Hall conductivity values, between spin ice and Kagome ice states, extracted from Fig. 2(b).

## 10. Study long range ordering and short-range ordering at Ir subslattices via the Resonant X-ray scattering:

The energy of incident x-ray is tuned near the Ir L absorption edge. To study the antiferromagnetic (AFM) type of ordering, (002) reflection which is forbidden by Bragg's Law was chosen. Fig. S7(a) summarizes the energy dependence of (0 0 2) peak at 5K,25K and 65K. There is no abrupt change in intensity as a function of temperature. Similar behavior in the l scan as shown in Fig. S7(b). No AFM magnetic ordering has been observed. To study the ferromagnetic (FM) type of ordering, (111) reflection which is allowed by Bragg's Law was chosen. Fig. S7(c) and Fig.S7(d) describe the energy dependence and l scan of (1 1 1) peak at 5K,25K and 65K, respectively. There is no change in peak intensity for neither of them, suggesting that the peak signal observed here is mainly contributed from the Bragg reflection, rather than FM magnetic ordering.

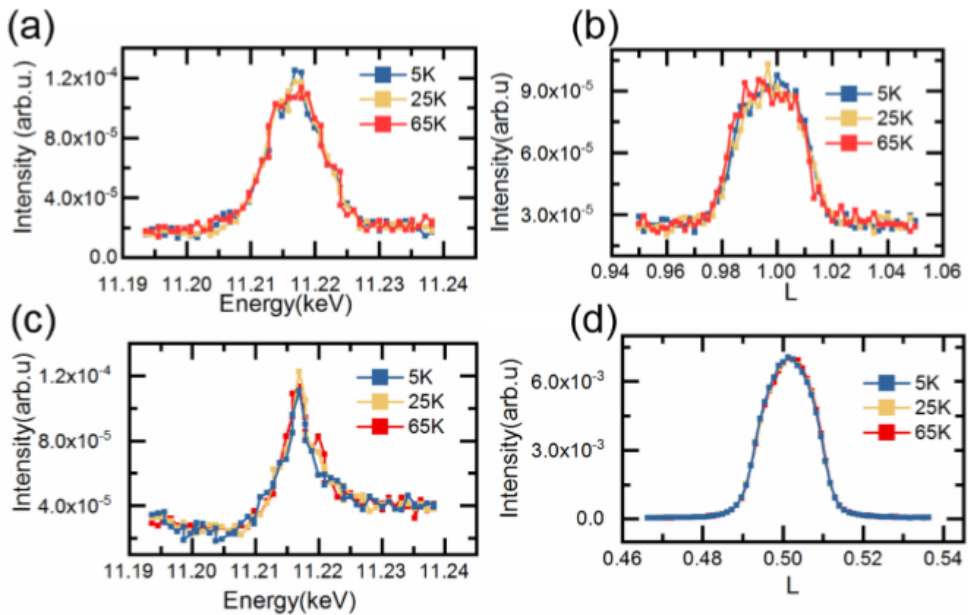

FIG S7: Resonant X-ray scattering with incident X-ray tuned near the Ir L edge. (a) and (b) summarizes the energy dependence and l scan of (0 0 2) peak at 5K,25K and 65K, respectively. (c) and (d) summarizes the energy dependence and l scan of (111) peak at 5K,25K and 65K, respectively.

To study the short-range ordering at the Ir sublattices, mapping in the reciprocal space has been performed below and above the onset temperature of spontaneous Hall effect with the incident X-ray energy tuned near the Ir L edge. Results are summarized in Fig. S8(a)-(c). In each part of the figures, the left column describes the measured reciprocal space at 5K. The middle panel describes the measured reciprocal space at 65K. The right panel describes the intensity difference across the measured area between 2 different temperatures. To be specific, the difference is calculated by subtracting the results of 5K from the results of 65K. The top and bottom row in each figure represent the projection of the reciprocal space mapping in **HK** plane and **HL** plane, respectively. Since there is no observed clear change in the peak intensity, the results suggest no observation of short-range ordering at the Ir sublattices.

(a)

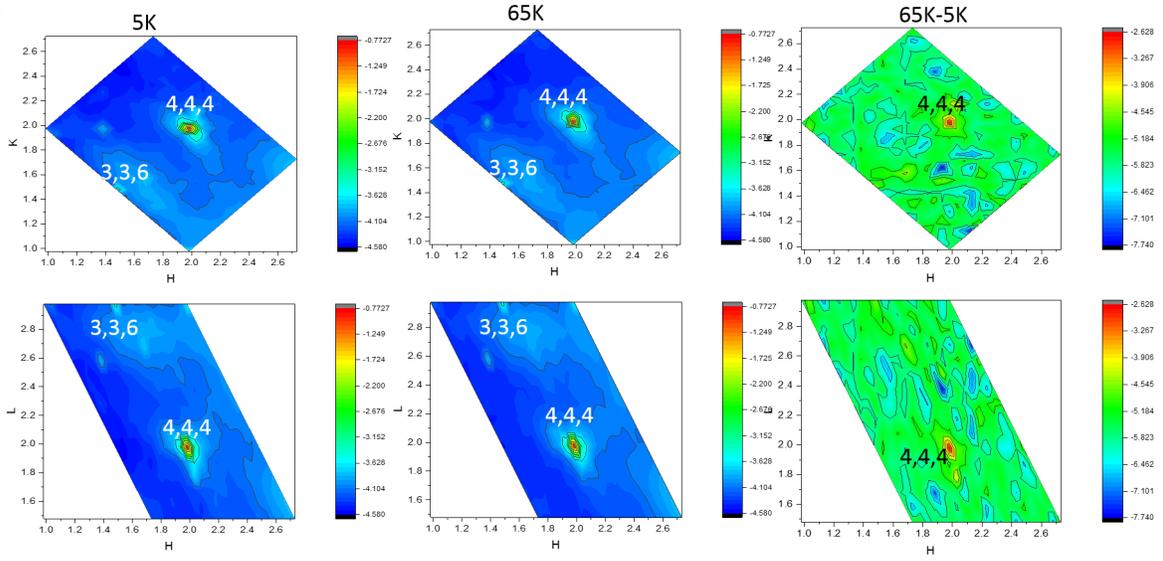

(b)

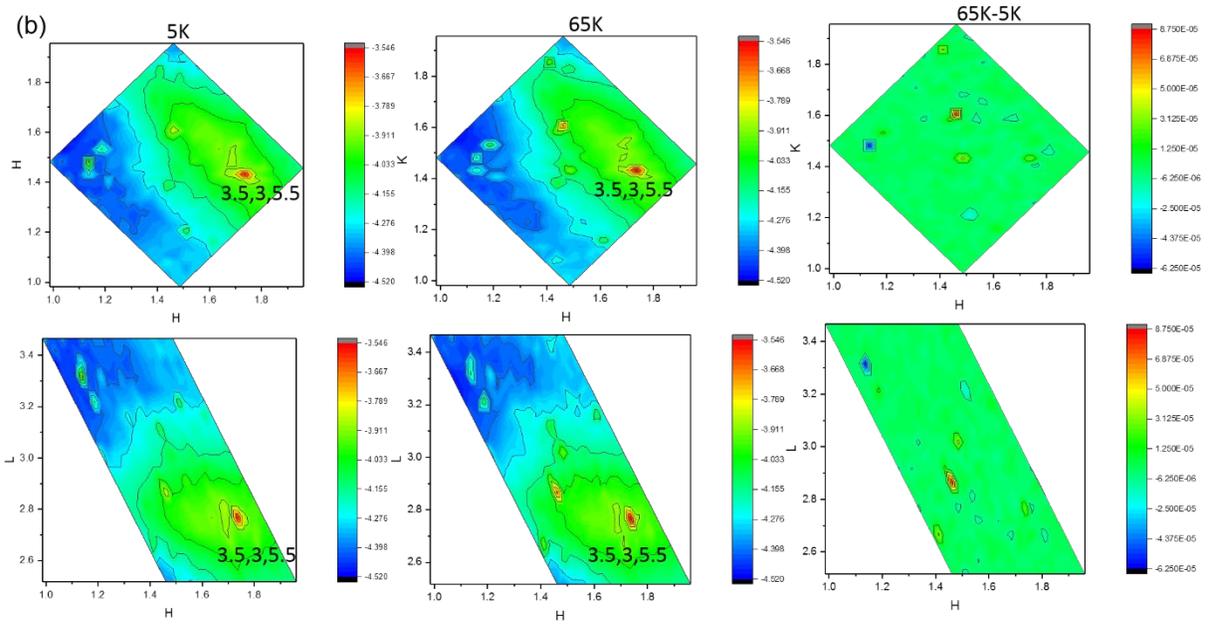

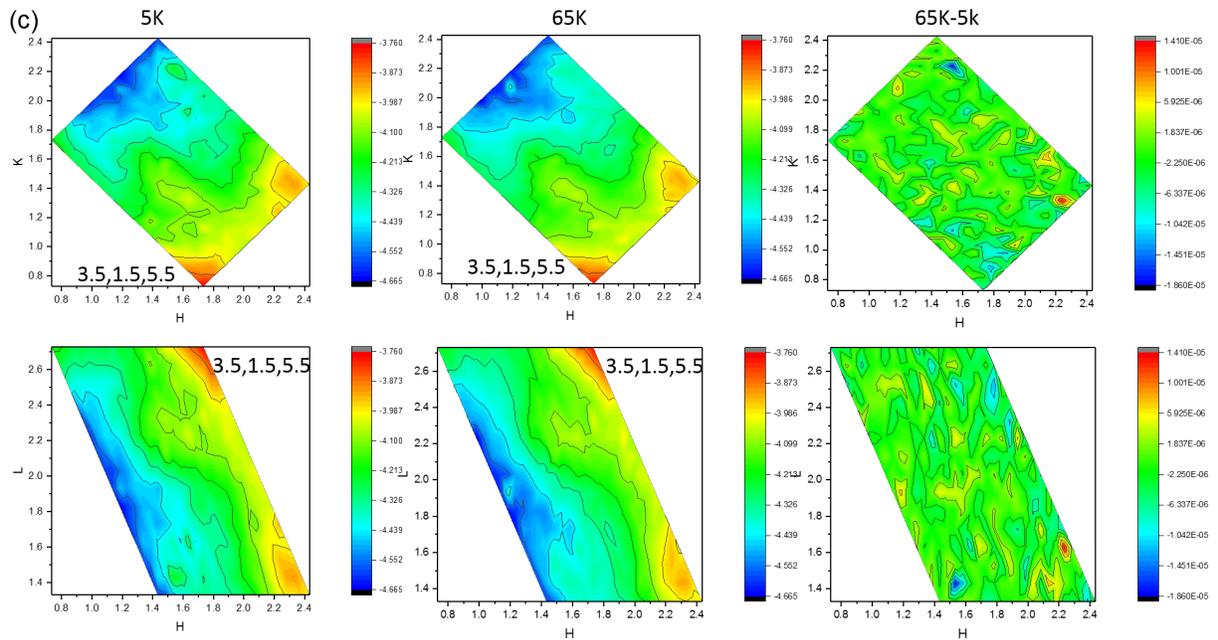

FIG. S8: (a)-(c) describe the reciprocal space mapping across the onset temperature of spontaneous Hall effect around (444), (336) and (3.5,3.5,3.5) reflection peak, respectively ,with the incident X-ray energy tuned near the Ir L edge.